\documentclass[times, twoside]{zHenriquesLab-Style}
\usepackage{xcolor} 
\definecolor{LightGray}{gray}{0.9}
\leadauthor{Ruscone} 

\begin{document}

\title{Building multiscale models with PhysiBoSS, an agent-based modeling tool}
\shorttitle{Building multiscale models with PhysiBoSS}

\author[1,2,3,\Letter]{Marco Ruscone}
\author[4,5]{Andrea Checcoli}
\author[6]{Randy Heiland}
\author[1,2,3]{Emmanuel Barillot}
\author[6]{Paul Macklin}
\author[1,2,3,*,\Letter]{Laurence Calzone}
\author[1,2,3,*,\Letter]{Vincent No\"el}

\affil[1]{Institut Curie, Université PSL, F-75005, Paris, France}
\affil[2]{INSERM, U900, F-75005, Paris, France}
\affil[3]{Mines ParisTech, Université PSL, F-75005, Paris, France}
\affil[4]{Centre de Recherche des Cordeliers, Sorbonne Université F-75005, Paris, France}
\affil[5]{INSERM, U1138, F-75005, Paris, France}
\affil[6]{Department of Intelligent Systems Engineering, Indiana University, Bloomington, IN, USA}
\affil[*]{Both authors contributed equally to this work.}

\maketitle

\begin{abstract}
Multiscale models provide a unique tool for studying complex processes that study events occurring at different scales across space and time. In the context of biological systems, such models can simulate mechanisms happening at the intracellular level such as signaling, and at the extracellular level where cells communicate and coordinate with other cells. They aim to understand the impact of genetic or environmental deregulation observed in complex diseases, describe the interplay between a pathological tissue and the immune system, and suggest strategies to revert the diseased phenotypes. The construction of these multiscale models remains a very complex task, including the choice of the components to consider, the level of details of the processes to simulate, or the fitting of the parameters to the data. One additional difficulty is the expert knowledge needed to program these models in languages such as C++ or Python, which may discourage the participation of non-experts. Simplifying this process through structured description formalisms --- coupled with a graphical interface --- is crucial in making modeling more accessible to the broader scientific community, as well as streamlining the process for advanced users. 
This article introduces three examples of multiscale models which rely on the framework PhysiBoSS, an add-on of PhysiCell that includes intracellular descriptions as continuous time Boolean models to the agent-based approach. The article demonstrates how to easily construct such models, relying on PhysiCell Studio, the PhysiCell Graphical User Interface. A step-by-step tutorial is provided as a Supplementary Material and all models are provided at: https://physiboss.github.io/tutorial/.
\end{abstract}


\begin{corrauthor}
vincent.noel@curie.fr, laurence.calzone@curie.fr, marco.ruscone@curie.fr
\end{corrauthor}

\section*{Introduction}
Multiscale modeling is a valuable tool in understanding complex biological systems, as it considers events occurring at various spatial and temporal scales. Such models are instrumental in investigating the interplay between intracellular level mechanisms, and intercellular interactions where cells communicate and coordinate. This is especially pertinent in the context of cancer, where multiscale models can be useful when studying the cross-talk between the microenvironment components, offering insights into the mechanisms of disease progression and potential therapeutic strategies.

In this context, we developed hybrid models, which result in a broader representation of biological systems, blending discrete agent-based techniques with continuous mathematical models \citep{metzcar2019abmreview}. This approach allows for a detailed depiction of individual cell behaviors while simultaneously capturing the broader, continuous dynamics of the biological environment. Such models are instrumental in accurately simulating the intricate interactions within cancerous tissues, shedding light on the complex interplay of cellular and molecular factors. However, developing these models can be challenging, often requiring proficiency in programming languages like C++ or Python, which might not be accessible to all researchers. 

An important advancement in this field was the introduction of PhysiBoSS\citep{letort_physiboss_2019,ponce-de-leon_physiboss_2022}. This add-on to PhysiCell\citep{ghaffarizadeh_physicell_2018} enhances the modeling process by integrating intracellular descriptions into the agent-based approach. PhysiBoSS utilizes MaBoSS\citep{stoll_continuous_2012,stoll_maboss_2017}, a tool that models signaling pathways as Boolean networks, thus simplifying the description of intracellular models. In addition, PhysiCell Studio\cite{physicell_studio_2023}, a graphical interface compatible with both PhysiCell and PhysiBoSS, further streamlines model development, catering to users with varying programming expertise. Despite these advancements, building a model can still prove complex for non-computational researchers who approach the software for the first time. However, we have streamlined the process significantly, making it more accessible and user-friendly. The models presented in this paper showcase a range of complexities and features, each highlighting a different aspect of multiscale modeling challenges and their solutions. This paper aims to demonstrate the construction of such multiscale models to answer biological questions, and guides the readers through the practical implementation of these models, demonstrating their utility in cancer research.

\section*{Methods}
\subsection*{Agent-based modeling with PhysiCell}
Agent-based modeling relies on a computational approach that uses autonomous, interacting software agents to study the behaviors of a system. An agent represents a single individual with its own state and behaviors that can react to other agents or the surrounding environment. Agent-based models allow for studying the emergence of complex population events from a simple set of agents’ behaviors. 
In medical science, an agent can represent a cell that can interact with other cells or its microenvironment. With this approach, it is possible to simulate different biological scenarios, study collective cellular behaviors, and test hypotheses {\textit{in silico}}. 
At present, many agent-based frameworks are available, with different characteristics to better answer different modeling needs \citep{metzcar2019abmreview}.
In this context, the C++ PhysiCell framework uses a center-based approach, simulating mechanical and phenotypical cell dynamics, as well as the diffusion of substrates to represent cellular respiration, paracrine communication, and more \cite{ghaffarizadeh_physicell_2018}. 
PhysiCell enables the customization of the simulations through a general configuration XML file, with optional specifications of initial cell positions and cell rules in CSV files. More recently, it was extended with a \emph{modeling grammar} that connects signals (e.g., diffusing chemical factors) with changes in cell behaviors, to help users straightforwardly model the stimuli perceived by an agent and its behavioral reactions \cite{johnson2023digitize}. PhysiCell includes dictionaries of available signals and behaviors for use in PhysiBoSS models, and PhysiCell Studio can use these pre-populated dictionaries to graphically construct model rules.

\subsection*{Logical modeling with MaBoSS}

Logical modeling provides an efficient way to study and represent complex behavioral patterns in biology. This method involves representing biological entities, such as genes, proteins, or full pathways, as nodes within a network. Using a Boolean approach, each node is a variable of the model that can take two values, 0 for absent or inactive and 1 for present or active, and the update of these variables is monitored by logical rules that link all the inputs of a node with the logical connectors OR, AND, and NOT. This type of models can be used to explore patients' responses by simulating various initial conditions and accounting for mutations observed in patients by forcing the values of the corresponding variables in the model.
MaBoSS is a C++ software package for simulating Boolean models using continuous time Markov processes \citep{stoll_continuous_2012, stoll_maboss_2017}. It applies an asynchronous update scheme, which allows the description of heterogeneous responses. By associating transition rates to each variable, for both activation and inactivation, it generates continuous trajectories with a notion of physical time. 
MaBoSS uses two files for describing the model: the BND file which contains the information about the Boolean network, and the CFG file which contains the simulation settings.

\subsection*{PhysiBoSS framework}

PhysiBoSS is an add-on of PhysiCell that integrates a MaBoSS engine inside each agent. 
This approach adds a new layer of description of the cell, with a specific Boolean model that represents the cell's intracellular signaling dynamics. The Boolean network can be the same for all the cells or separate networks can be assigned to each cell type.
At each simulation step, the agent (cell) can collect different stimuli that modify the activity of some specific nodes of the network (input nodes). Next, the MaBoSS engine computes the model trajectory that can cause the switch of the so-called phenotypic nodes (or output nodes). Those nodes can then trigger some specific cell actions (motility, secretion, uptake, death, etc.).
PhysiBoSS uses as input the same configuration files of PhysiCell, and the BND and CFG MaBoSS input files.

\subsection*{Mapping agent-based to intracellular models}
PhysiCell provides a dictionary of signals and a dictionary for behaviors, aimed at giving better accessibility to all the signals perceived by each agent and all possible behaviors that an agent can express. PhysiBoSS uses these data structures to simplify the connection between PhysiCell and MaBoSS, giving access to the PhysiCell/MaBoSS mapping through the configuration file and so, drastically diminishing the amount of C++ code necessary to develop a model. Mapping can be of two types: (1) input mapping, which links a PhysiCell signal to a MaBoSS (input) node by using activation thresholds, or (2) output mapping, which links a MaBoSS (output) node to a PhysiCell behavior by using values representing the Boolean state. Implementation details about the mapping are available in the supplementary section \textbf{S1.2}. 

\subsection*{Time synchronisation}
The intracellular model is updated periodically, according to the value of  {\texttt{intracellular\_dt}}. The {\texttt{scaling}} parameter is also available to match the time scale of the intracellular model to the time scale of the agent-based model. Finally, to account for biological phenomena such as cellular desynchronization, an option is available for stochastic update time. More information about the implementation of time in PhysiBoSS is available in the supplementary sections \textbf{S1.3}  and \textbf{S1.4}.


\section*{Results}
PhysiBoSS performs simulations of models that combine intracellular molecular description (with MaBoSS) and physical intercellular communication (with PhysiCell). With this approach, it is possible to study the impact of events that occur inside the cell at the level of the population and the effect a treatment may have considering physical features.  

We present three examples of multiscale models: (1) a modified version of a previously published model of cell fate decision processes in response to death receptor engagement and the effect of a TNF treatment on these decisions, (2) a cell cycle model for investigating the consequences of genetic perturbations in signaling, and (3) a simplified model of immune cell differentiation. For each of these models, we provide step-by-step procedures as supplementary materials to build these models, which can be used as templates for any other project. In the text below, we present and analyze the expected behaviors for each of the three sample models to serve as a reference for self-learners as they work through the article and demonstrate the range of integrated model types that can readily be built. The materials also include an additional improved version of a model of cell invasion already published \cite{ruscone2023multiscale}.

\subsection*{Cell fate model upon TNF treatment}
Upon cell death receptor engagement, different phenotypes can be triggered depending on the status of some cell components. Programmed cell death, through necroptosis or apoptosis, or survival through the NF-$\kappa$B pathway can be activated. A previously published Boolean model of the complex intertwined networks leading to these cell fates was used \cite{calzone_mathematical_2010} and integrated into PhysiBoSS \cite{letort_physiboss_2019} to study the effect of a TNF treatment on a population of interacting cells by varying the type of treatments (continuous vs. pulsating) and the composition of the population (to explore the efficacy of the treatment of a heterogeneous population). The model presented here is an improved version of the initially published one modified to fit the evolution of the tool. 

\subsubsection*{Analysis of the intracellular model}
The intracellular model considers two receptors, Fas and TNF, and studies the conditions that lead to either survival ({\it Survival}), programmed cell death ({\it Non\_apoptotic\_Cell\_Death} or {\it NonACD}), or apoptosis {(\it Apoptosis}) (see Supplementary Materials, figure \textbf{S9}). With MaBoSS\cite{stoll_maboss_2017}, the proportion of the three cell fates can be quantified and differences appear with varying initial conditions or types of treatments: upon continuous TNF receptor activation, most of the cells (95\%) will trigger apoptosis, while a small population of cells will activate either necroptosis (referred to as non-apoptotic cell death or NonACD) (3\%) or NF-$\kappa$B-driven survival (2\%); when cells are treated in a pulsating manner (every 40 hours for 20 hours), the simulation of a population of individual non-interacting cells shows very little difference, even though, in contrast with the continuous treatment, at time 100, all cells have undergone apoptosis (Figure \textbf{S11}). 

This model can also simulate gene mutations and the impact they have on the cell fate distribution. For example, the double mutant {\textit{IKK++}}/{\textit{cFLIP++}} shows a shift of phenotypes following TNF treatment to only obtain resistant cells, with NF-$\kappa$B fully active. 

\subsubsection*{Integration of the Boolean model in PhysiBoSS}
When integrating a Boolean model into PhysiBoSS, there are several aspects to consider: (1) the time scales of the two models which may require synchronization between the two scales, and (2) the connection between the Boolean intracellular model and the agent-based model. 

\begin{figure*}[ht]
\begin{center}\includegraphics[width=0.80\linewidth]{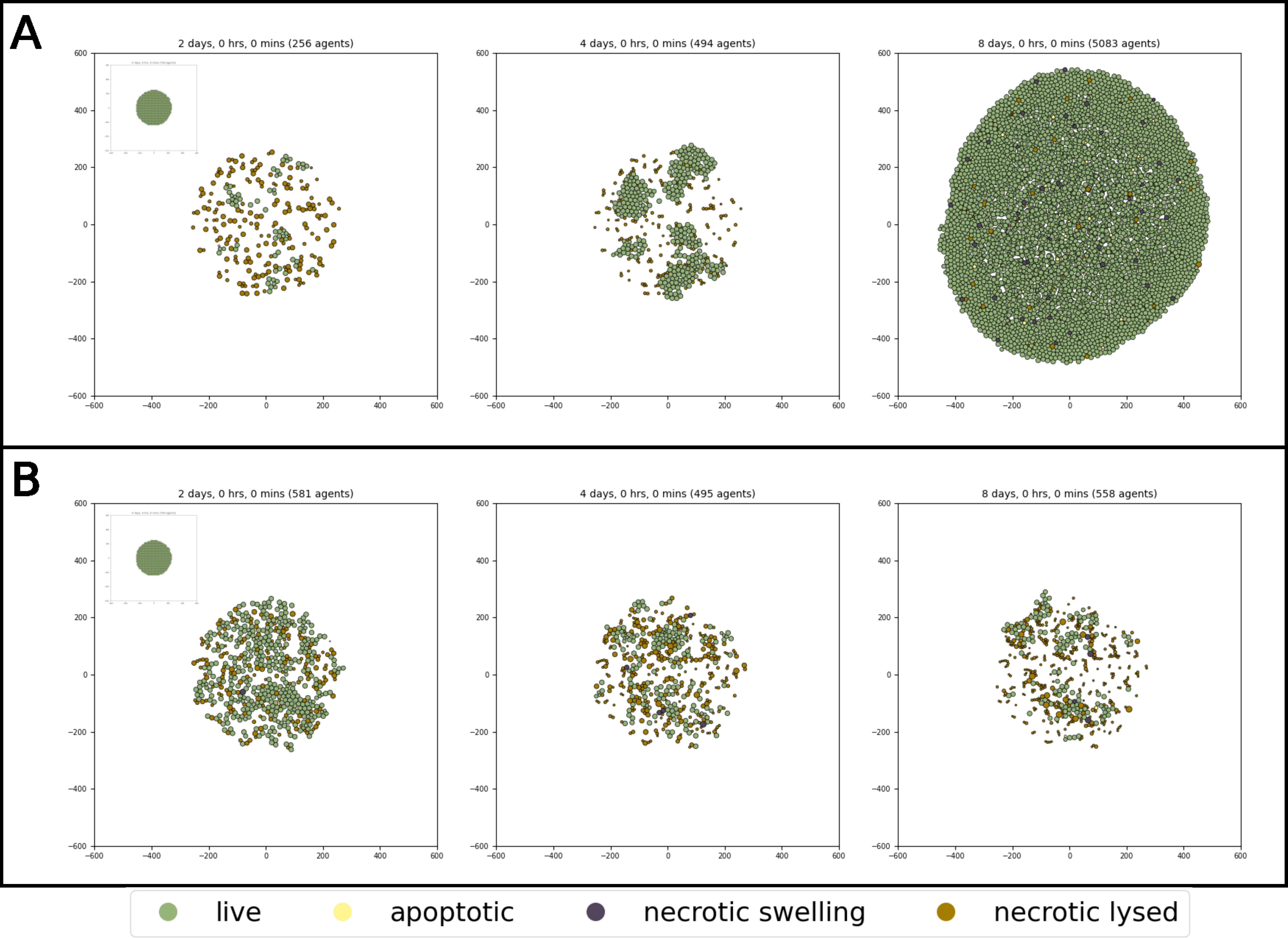}\end{center}
\caption{Simulation of the cell fate model upon TNF treatment, at t=2,4,8 days. A) MaBoSS simulation of a prolonged and continuous TNF treatment. B) MaBoSS simulation of pulses of TNF treatment.}
\label{fig:tnf_simulation}
\end{figure*}

The synchronization of the two time scales is a difficult task as intracellular and extracellular events may not have the same scales. The two parameters controlling timing are: {\texttt{scaling}} and {\texttt{intracellular\_dt}}. Since the standard PhysiCell simulation time unit is in minutes, while the cell fate model's unit is in hours, the {\texttt{scaling}} parameter needs to be set to 60, thus converting the MaBoSS model unit to minutes. The second parameter specifies how often the cell agents should execute and update their MaBoSS models. In this specific case, the asymptotic behavior of the system is considered, which is reached after 24h, setting the parameter {\texttt{intracellular\_dt}} to 1440 min (24h = 1440min). To avoid having all our cells respond in synchrony to the TNF treatment, we set the value of {\texttt{time\_stochasticity}}, a parameter responsible for producing slightly different periodic updates, to 0.5 (this parameter describes the deviation of the distribution, and is explained in the supplementary section \textbf{S1.4}).
The next step consists in the mapping of the two models, described by three rules. The first mapping rule is an input rule, which describes the condition in which the TNF ligand in the vicinity of the cell will be able to activate the TNF input node of the intracellular model, then triggering downstream intracellular events. The next two mapping rules are output rules, connecting the intracellular phenotypes to behaviors of the PhysiCell simulation. In the cell fate model, there are three outputs, two of which correspond to the two death phenotypes. The first output mapping rule will link the {\it Apoptosis} node to the {\texttt{Apoptosis}} behavior, which in PhysiCell is controlled by a fixed activation rate. To be uniquely controlled by the {\it{Apoptosis}} node, we set this rate to 0 when the node is inactive, and to a very high value (1e+6) when the node is active (thus ensuring the apoptosis is deterministically activated at the next PhysiCell time step). The second output mapping rule rule is similar for the necrosis node ({\it NonACD}) which is linked to the activation rate controlling the PhysiCell {\texttt{Necrosis}} behavior. Finally, the last phenotype, {\it Survival} is left without any mapping, as it represents the complement of the two death phenotypes, so it can be described as a resistant phenotype to the TNF treatment. 
Note that variations on this cell fate model exist where the NFkB pathway is linked to an autocrine secretion of TNF, which could create a feedback loop in our model\cite{stoll2022upmaboss}. A brief description of how to create this behavior is described in the supplementary, section \textbf{S3.8}.
To simulate the TNF treatment in time, a function was added, controlled by user parameters. Note that PhysiPKPD \cite{bergman2022physipkpd}, a recent add-on of PhysiCell, also facilitates the simulation of many types of treatment.
For the prolonged TNF treatment, the parameter {\tt{treatment\_duration}} was set to 11520 minutes (8 days), more than our simulation maximal time. In Figure~\ref{fig:tnf_simulation}A, it can be observed that, while most of the population is killed either by apoptosis or necrosis on day 2, a resistant population emerges and leads to a large proliferating population on day 8.

To reproduce the effect of a pulsatile treatment, the parameters {\texttt{treatment\_duration}} and {\texttt{treatment\_period}} were modified to simulate a treatment of 2000 minutes happening every 3440 minutes. In Figure~\ref{fig:tnf_simulation}B, the size of the population of tumor cells decreases after each treatment. Such treatments---if clinically validated---could potentially be used to prevent the formation of a population resisting the TNF treatment, as well as to reduce the toxicity of the treatment. 

Finally, to explore more functionalities of PhysiBoSS, we also produced a version of the model accounting for the observed necrotic core of the tumor due to the lack of oxygen (see supplementary materials, section \textbf{S3.6}), and describing the impact of \textit{IKK++} - \textit{cFLIP++} double mutations on the outcome of the treatment (supplementary materials, section \textbf{S3.7}).
By building this model from the original PhysiBoSS into PhysiBoSS 2.2, we showed that only a few simple steps are now needed, allowing a much wider user base to build complex models easily. A complete description of the steps necessary to build this model is available in the supplementary, section \textbf{S3}. 
\subsection*{Boolean cell cycle model}
The cell cycle is a complex system, controlled by cyclins and cyclin-dependent kinases (CDKs) which act as checkpoints to ensure that the necessary steps are performed and the cycle can progress. The loss of control in proliferation is one of the hallmarks of cancer, which may be due to some alterations in the signaling pathways that lead to the transcription of cell cycle genes.
PhysiCell however represents this cycle as a straightforward process, where each phase has a fixed transition rate, and no signaling is involved to perturb it. With this example, we wanted to integrate with PhysiBoSS a more realistic cell cycle model and show how we can reproduce the effect of known mutations. To this end, we used a published Boolean model of the cell cycle from Sizek et al.\cite{sizek_boolean_2019} as an intracellular model, and linked it to the transitions between the different phases to control the progression of the PhysiCell cell cycle.

\begin{figure*}[ht!]
\begin{center}\includegraphics[width=0.80\linewidth]{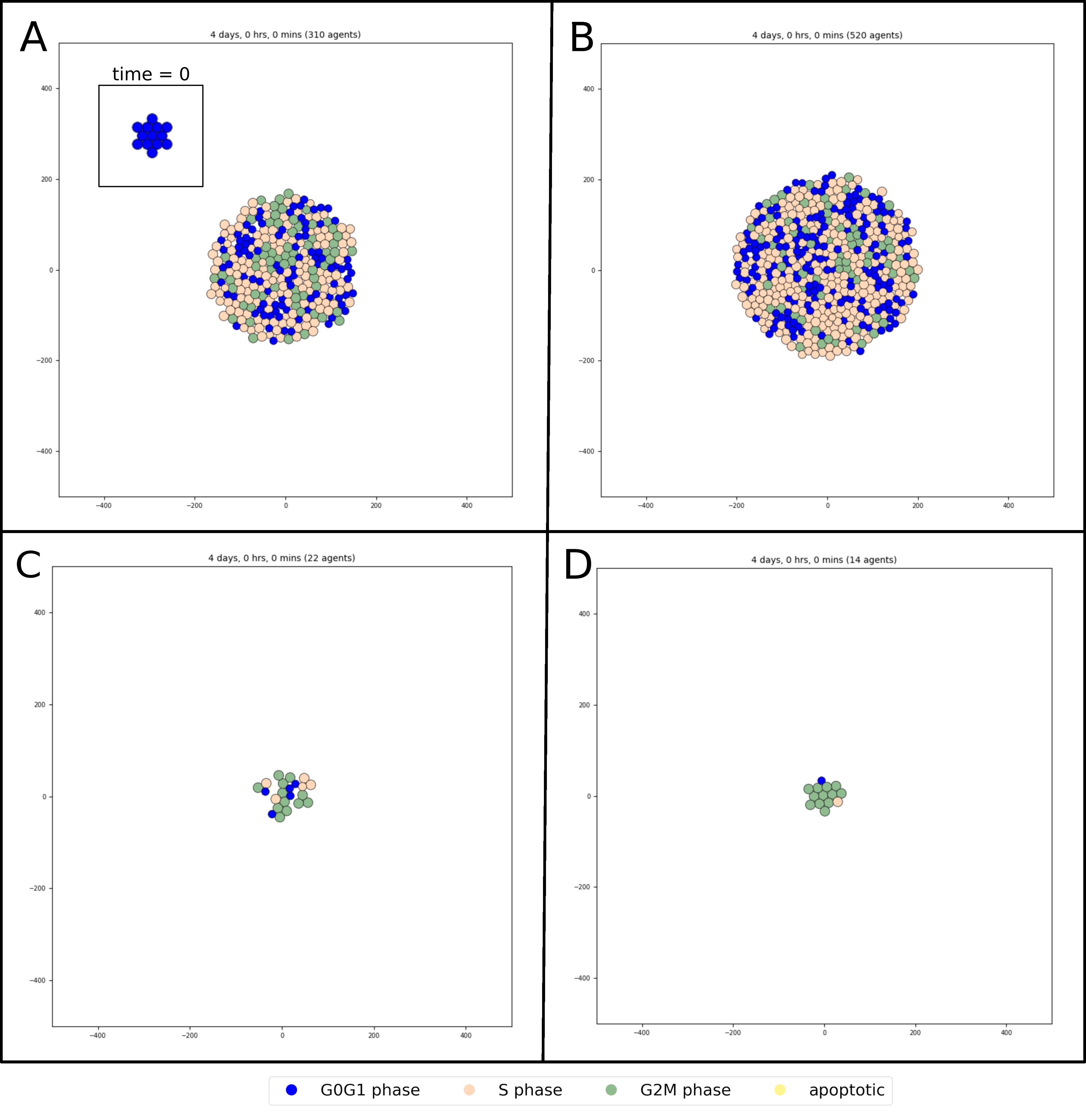}\end{center}
\caption{Simulation of the Sizek cell cycle model. A) Wild-type simulation at both time 0 and after 96 simulated hours. B) Knock-in of p110 inactivates the apoptosis pathway which increases the growth rate of the population, with 520 cells after 96 hours vs. 310 cells in Wild Type condition. C) FoxO3 knock-out simulation slows down the cell cycle, diminishing the number of cell divisions, with 22 cells after 48 simulated hours. D) Plk1 knock-out simulation causes the majority of cells to be stuck in G2/M phase. All the simulations were executed with a value of \textit{scaling} of 37.5 and \textit{intracellular\_dt} of 2.5}
\label{fig:cell_cycle}
\end{figure*}

\subsubsection*{Analysis of the model}
In their work, Sizek and colleagues built a Boolean model that reproduces the cell cycle progression, including apoptosis and growth signals. The model is composed of 87 nodes and captures PI3K/AKT1 activity during the cell cycle and its role in the deregulation of PLK1 and FOXO3. The perturbations can lead to different cell fates such as G2 arrest characterized by a sustained activity of Cyclin B, or mitotic catastrophe caused by Casp2 activation during mitosis. 
To integrate the Boolean network into PhysiBoSS, we first performed some analysis of the model to decide which nodes would be responsible for the switch between cell cycle phases.
The analysis was done using MaBoSS \cite{stoll_maboss_2017} and included in a Jupyter Notebook where we simulated the wild-type model with different initial conditions, and mutants (see supplementary materials, Cell\_cycle\_boolean\_analysis.pdf). 

The model analysis shows an interplay between components of the cell cycle and the apoptotic pathway, highlighting the role of Casp3, a read-out of cell death, which spontaneously and gradually gets activated after several cycles. The model can reproduce the sequential activation of the cyclins: Cyclin E, Cyclin A, and Cyclin B, and their oscillation until Casp3 gets fully activated. However, we observed that this sequence is not always preserved and can lead to an incomplete cell cycle, such as Cyclin E and Cyclin A activation not followed by a Cyclin B activation. An analysis of the transitions between phases is also available in the Jupyter notebook mentioned above. 

The initial model reported published mutations and reproduced their phenotypes, which were then confirmed with the MaBoSS simulations (see \texttt{Cell\_Cycle\_Analysis} notebook in supplementary). Among these mutations, we focused on the role of PLK1, FOXO3, p110, and PI3K. The loss function of PLK1 (\textit{PLK1} node is set to 0) leads to an overactivation of Cyclin B, indicating that the cells may be stuck in the G2 phase, with no observed apoptosis. A knock-out of FOXO3 (\textit{Foxo3} node is set to 0) leads to a failure of cytokinesis. In this condition, most cells are unable to separate the cytoplasm and to complete division. Some will start apoptosis, while the majority of them will stay in this failed state (characterized by none of the cyclins nodes being active).
Finally, the knock-in mutation of p110 ({\textit{p110}} node is set to 1) shows an increase in the activity of AKT leading to a decrease in the activity of the apoptosis pathway. 

Among the in-built cell cycle models proposed by PhysiCell, we selected one of the simplest, the Flow Cytometry model, composed of 3 phases and 3 rates. In this model, a cell starts at the default phase "G0G1" and enters the cell cycle with a rate r01 to reach the "S" phase. From the "S" phase, it moves to the "G2M" phase with a rate r12. Finally, the cell divides and returns to the "G0G1" phase at a rate r20. 

With PhysiBoSS, it is possible to associate the transition rates of a cell cycle phase, to the state of a node of the Boolean model. To facilitate this pairing, we included in the Sizek model three phenotypic nodes that match the three transitions of the Flow Cytometry model: {\textit{G0G1\_entry}}, {\textit{S\_entry}} and {\textit{G2M\_entry}}.
The state of these nodes is determined by the activity of one or more Cyclins: CyclinD1 and CyclinA control the {\textit{G0G1\_entry}}, CyclinA and CyclinE control the {\textit{S\_entry}} and finally CyclinB controls the {\textit{G2M\_entry}}. 
The introduction of these three read-out nodes does not affect the behavior of the network but provides a single Boolean node for each transition between the three phases.

\subsubsection*{Integration of the Boolean model in PhysiBoSS}

To include the Sizek model in PhysiBoSS, we focus on the two parameters that control the time synchronization and the mapping.
To synchronize the time between the two models, we started by considering a cell cycle duration of 24 hours. Since a full cell cycle in MaBoSS is achieved in 24 units of time, we proceeded to set the scaling value to 60, similar to what was done with the previous TNF model. However, this choice did not result in a 24-hour cell cycle, but a longer one of 39 hours, partially due to the incomplete cycles mentioned in the previous section. To fix this, we calculated a correction for the scaling factor, setting it to 40 and reproducing the expected cellular behaviors (see supplementary materials, section \textbf{S4.2}). The time interval is set to a small value ({\texttt{intracellular\_dt} = 1 min) since, contrary to the previous model, here it is important in this model to capture transient effects.
The model does not take into account environmental conditions, making irrelevant the mapping of input nodes. However, it is possible to specify in the intracellular configuration the initial state of the inputs of the model, such as the node \textit{Trail} (death signal) or \textit{GF} (growth factor).
We proceeded to connect the previously defined phenotype nodes to the corresponding behaviors, associated with the controls of the cell cycle transition rates. The  \textit{S\_entry} node is connected to the behavior \texttt{Cycle entry}, \textit{G2M\_entry} to \texttt{exit from cycle phase 1},\textit{ G0G1\_entry} to \texttt{exit from cycle phase 2}. Finally, the node \textit{Casp3} is connected to the behavior \texttt{apoptosis} which concretely modifies the rate of activation of the apoptotic death model.
The basal value of all the rates is set to 0. When one of the nodes regulating the phenotype is activated, the transition rate is fixed to a very high value (1e+6) to immediately trigger the phase switch or the apoptotic death. When the node is inhibited, it restores the basal value of the transition rate. 

The initial population of the PhysiBoSS simulations are set to 13 cells (agents) growing to 310 in 96 hours (Figure\ref{fig:cell_cycle}A). The phases follow a proper order in individual cells, but not all cells are in the same phase of the cycle as expected in a desynchronized population of cells. 

We further tested the impact of mutations at the population level, by selecting the appropriate node to mutate and assigning it a value of 0 (knock-out) or 1 (knock-in). 
The mutant \textit{p110} overexpressed (\textit{p110} nodes fixed to 1) results in decreasing the apoptosis with a consequent increase of the proliferation rate, bringing the final number of cells after 96 hours from 13 to about 520 (Figure~\ref{fig:cell_cycle}B).
Next, we tested \textit{FoxO3} knock-out (\textit{Foxo3} node fixed to 0). The simulations show that the cells go through one cell cycle before either dying or slowing down the proliferation. The cells are not arrested in a specific phase of the cycle, but they keep proliferating at a very low rate (Figure~\ref{fig:cell_cycle}C). Finally, \textit{Plk1} knock-out (\textit{Plk1} node fixed to 0), as expected from the MaBoSS analysis, causes the majority of the cells to get stuck in G2/M phase, in a cell cycle arrest (Figure~\ref{fig:cell_cycle}D). 

In conclusion, the multiscale model of a detailed molecular description of the cell cycle reproduces the complexity of the cell cycle at the single and multicellular level, allowing not only the modification of the duration of the cell cycle but also the realization of mutations and the exploration multiple initial conditions (corresponding different extracellular contexts). Some phenotypes were not observable with the intracellular model only, such as the slowing down of the cycles, but could be observed with the PhysiBoSS model. A complete description of the steps necessary to build this model is available in the supplementary, section \textbf{S4}. 
\subsection*{Immune cell differentiation}
The examples previously presented assumed that all cells were of the same type. With PhysiBoSS, it is possible to consider interactions among several cell types with different intracellular models. In this example, we showcase a simple model of cell differentiation, where a cell of a specific type can transition into a different, user-defined cell type.
Moreover, we demonstrate how different signals (diffusible chemical factors, type-specific contacts) can be used as inputs to regulate key cell behaviors.
The model encompasses six different cell types and relies on two different Boolean models.\\ 

\begin{figure*}[ht!]
\begin{center}\includegraphics[width=0.80\linewidth]{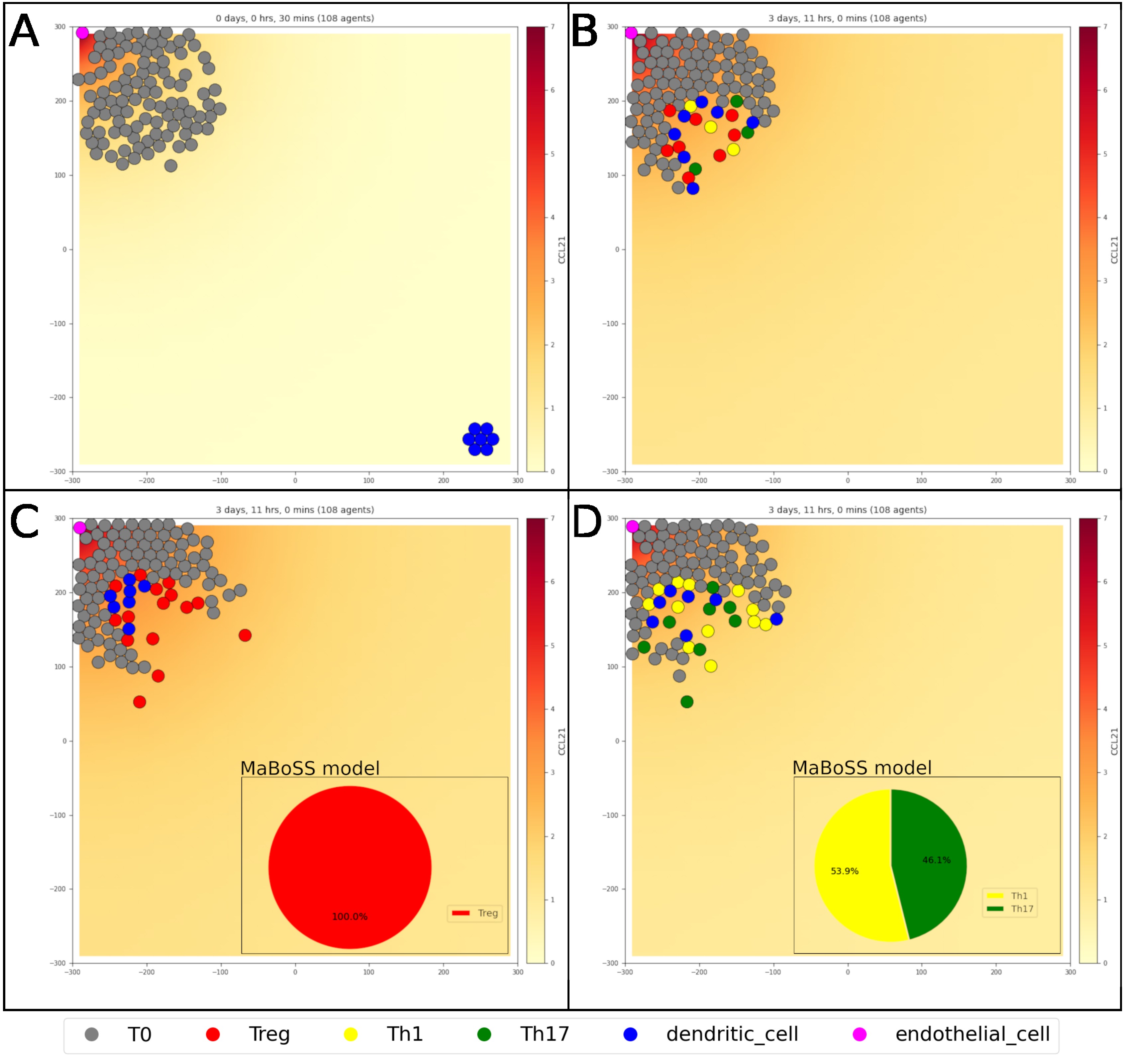}\end{center}
\caption{Simulation of the T cell differentiation model in 2 and 3 dimensions. A) Initial population of T cell (gray), with an endothelial cell (pink) secreting CC21. A population of dendritic cells (blue) is attracted towards the source of CCL21. B) Upon contact, the dendritic cells trigger the receptors of the naive T cell, which start the differentiation process according to the outputs of the intracellular model, into Treg (red), Th1 (yellow), and Th17 (green). C) Simulation of the T cell differentiation with NFkB knock-out, resulting in only Treg. D) Simulation of the T cell differentiation with FOXP3 knock-out, resulting in only Th1 and Th17.}
\label{fig:differentiation}
\end{figure*}

\subsubsection*{Analysis of the intracellular models}
The Boolean model for cell differentiation is adapted from a previously published model of Corral-Jara and colleagues \cite{corral-jara_interplay_2021}, which describes the processes of T cell differentiation. 
The model is based on experiments performed on naive CD4+ T cells (referred to as T0), which depending upon the effect of external stimuli, can differentiate into either a Type 1 helper cell (Th1), a T helper 17 cell (Th17), or a regulatory T cell (Treg).
Note that Corral-Jara's model has been designed in GINsim software in such a way that some nodes are multi-valued to represent different levels of activation. MaBoSS does not allow discrete levels and all multi-valued nodes are Booleanized into two variables, e.g., MHCCII has been split into two variables \textit{MHCII\_b1}, \textit{MHCII\_b2}.
We also use a simple phenomenological model for dendritic cells with a small set of nodes to describe their behavior.  
The model (Figure \textbf{S24}) encompasses a total of 4 nodes, of which 3 inputs (Maturation, Contact, CCL21) and 1 phenotype node (Migration). A more complex model can later replace this simple model.\\
In this model, under the chemoattractant effect of the CC motif chemokine ligand 21 (CCL21), a cytokine constitutively expressed in secondary lymphoid organs (such as lymph nodes), a population of mature dendritic cells (mDCs) is attracted towards the draining lymph node. Concretely, the activation of the node {\textit{CCL21)}, when the node {\textit{Maturation}} is already active, activates the node {\textit{Migration}}.\\
mDCs express a set of ligands capable of triggering the differentiation of the T0 cell population into 3 different subsets of CD4+ T cells, once in the lymph nodes.
Among these ligands, we can cite Interleukin-12 (IL-12), Interleukin-1β (IL-1β), and other cytokines as Interleukin-6 (IL-6) or Transforming growth factor beta (TGF-β). We chose not to include all those nodes in our mDC model and just represent them by a single node {\textit{Contact}}. The activation of this node turns off the migration of the mDCs.
Within the Corral-Jara's MaBoSS model, these ligands' nodes are already present as input nodes.
The CD4 + T cell model includes 3 master transcription factors considered as markers of differentiated T cells: RORgt (Th17), FOXP3 (Treg) and Tbet (Th1). Based on these nodes, we built three phenotype nodes (\textit{Th1}, \textit{Treg}, and \textit{Th17}), used later as output nodes, to better represent the different cell types. The relative logical equations have been constructed to avoid overlap between phenotypes so that each one is mutually exclusive. 

The model has been tested for mutants, to search for possible targets that can influence the probability of differentiation for the three cell types. Those mutants, introduced at the beginning of the simulation, should not trigger immediately the differentiation of the T0 cell, but rather have an impact on the differentiation process after contact with the mDC. Among the mutants, we found some cells differenting into Treg exclusively: (inhibitions of \textit{API}, \textit{NFKB}, \textit{LCK}, \textit{TCR}, \textit{RAS}, \textit{ITK}, \textit{ERK}, \textit{cFOS}, \textit{cJUN}, or \textit{IKK}), into a mix of Th1 and Th17: inhibition of \textit{IL1RAP}, \textit{IL1R1}, \textit{IL1R}, \textit{FOXP3\_2}, or activation of \textit{MINA}); and a mix of Treg and Th17 (inhibition \textit{STAT1}, \textit{Tbet}, or \textit{PLCG}). We also investigated the effect of modifying the activation rate parameters, to control more finely the proportions of Treg, and found that, for example, the activation rate of \textit{NFKB} can be lowered to increase the proportion of Treg, while the activation rate of \textit{FOXP3\_2} can be lowered to reduce their proportion (See supplementary materials, section \textbf{S5.4}).
\subsubsection*{Integration of the Boolean models in PhysiBoSS}
In this model, there are several different cell types: naive T cells (T0), dendritic cells, type 1 helper cells (Th1), T helper 17 cells, regulatory T cells, and finally lymphoid endothelial cells. For the integration of the two Boolean models presented above into PhysiBoSS, two intracellular models are created for the naive T Cells and the dendritic cells (see supplementary materials, Figure \textbf{S25} and \textbf{S26}). The other cell types are considered as agents with no intracellular description. As for the TNF example, the asymptotic behaviors of both the naive T cell and dendritic cell are considered. Based on the MaBoSS simulations, the two parameters, {\texttt{scaling}} and \texttt{intracellular\_dt}} are set to 1 (default value) and 6 (standard phenotype time step of PhysiCell), respectively. \\
For simplicity, we created one single endothelial cell secreting CCL21, located in an area representing the lymph node. We also created a population of T0 cells in the same area, as well as a distant population of dendritic cells (Figure~\ref{fig:differentiation}A). 
The initial state of the PhysiBoSS simulation assumes that the dendritic cells are mature, a condition in which they are expressing CCR7, a receptor that drives the migration of mature DCs (mDCs) towards secondary lymphoid structures (i.e., the lymph nodes).
We then created one input mapping in the dendritic cells, linking the substrate {\texttt{CCL21}} to the node {\textit{CCL21}}.
Upon activation of the {\textit{CCL21}} node within the DC network, mDCs move towards the source of CCL21, following its gradient combined with a random walk. 
Once in the lymph node, the DC moves with a random walk, as hypothesized in \cite{azarov2019role}. To achieve this, we used the rules' mechanism described in \cite{johnson2023digitize} and created a rule where the stochasticity of the chemotaxis is dependent on the quantity of CCL21, creating a saturating effect that progressively lowers the migration bias as the CCL21 quantity increases.

When mDCs and T0 cells are in contact, the differentiation process of naive T cells is triggered. The mDCs secrete major cytokines that are essential to mediate first the contact between DCs and T0 cells (a set consisting of IL-12, IL-1β, IL-6, TGF-β, and IL-23), and then to trigger the cascades leading to the three subsets of differentiated T cells, Th1, Th17, and Treg. 
For the sake of simplicity, instead of allowing each agent representing a dendritic cell to release cytokines, we encoded such interactions by activating the input corresponding to the cytokines within the T0 model. To do this, we created many input mappings that connect the contact of the dendritic cell with a T0 to the activation of the input nodes corresponding to the cytokines that are released by the dendritic cells.
In addition to the above-mentioned list of cytokines, input nodes triggered upon contact between dendritic cells and T0 include also: \textit{MHCII\_b1}, \textit{MHCII\_b2}, \textit{CD80}, \textit{CD4} and \textit{PIP2}.
The activation of these nodes is necessary to trigger the differentiation into Th1, Th17 or Treg (Figure~\ref{fig:differentiation}.B). To achieve this, we added three output mapping rules, linking the transformation into these cell types to the phenotype nodes {\textit{Th1}}, {\textit{Th17}}, {\textit{Treg}} presented previously. None of these new cell types have an intracellular model, so upon differentiation, they lose all the T0 properties. This choice was made to allow the implementation of specific behaviors for the different T cell types in future versions of the PhysiBoSS model.\\
We also included variants of this model representing the effect of two of the Th0 mutants described in the previous section: the knock-out of \textit{NFKB}, leading to a differentiation exclusively in Treg (Figure~\ref{fig:differentiation}.C), and the knock-out of \textit{FOXP3\_2}, leading to a complete absence of differentiation in Treg (Figure~\ref{fig:differentiation}.D). We also showed two other variants with lower activation rates of these two nodes (Figure \textbf{S27.B} and \textbf{D}), representing an incomplete inhibition. These examples showed how PhysiBoSS can describe the pharmacological control of T-cell differentiation. A complete description of the steps necessary to build this model is available in the supplementary, section \textbf{S5}. 

\section*{Discussion}
In this paper, we presented new functionalities of PhysiBoSS, which are drastically simplifying the process of creating models. We show that using the new mapping system, we can now easily connect the agent-based model to Boolean intracellular models. \\
While the previous version of PhysiBoSS required knowledge in C++ programming to allow the creation of models, with this new version the user can completely rely on PhysiCell Studio, the graphical interface of PhysiCell, to build a model from existing templates. These improvements are important both to the new users discovering the framework, and also to speed up the development of models by existing users. For some specific functionalities which still require writing code, such as the mechanisms regulating drug treatments, new add-ons of PhysiCell are being developed to simplify their accessibility. \\
While simple, we believe that the three models presented here cover enough functionalities to give a good overview of PhysiBoSS and provide broad examples to start from. We are providing in the supplementary materials a step-by-step guide for installing PhysiBoSS and PhysiCell Studio, and for building these models to allow newcomers to follow the process of creating them. 
The example of cancer invasion in the supplementary shows a better real-world example, and its comparison with the original models shows the simplicity and power of the mapping system. \\

Integrating biological data into these models is the next logical step for them to go beyond toy models. Different types of data can be used to address the different parts of the model: spatial data (spatial transcriptomic, multiplex immunofluorescence, ...) can be used for reproducing the disposition of the cell in the tissue \cite{johnson2023digitize}. 
Single-cell expression data can be used to infer cell-cell communication \cite{dimitrov2023liana}, and to personalize the intracellular model \cite{beal2019personalization}. Finally, many physical parameters could be obtained using time-lapse microscopy data.
However, tuning these parameters, even with appropriate datasets, would still be a difficult endeavor. In this article, we did not want to put too much emphasis on this, but it is a real challenge that may be addressed with machine learning approaches. New methods are needed in this field, and we believe the use of surrogate models \citep{preen2019towards, ROCHA2022115412} will prove itself fundamental. 
Despite these challenges, the improvements in PhysiBoSS presented here will facilitate the use of multiscale modeling and allow a larger community of users to apply these tools to their questions. 

\section*{Competing interests}
No competing interest is declared.

\section*{Author contributions statement}
M.R., V.N., R.H., L.C. developed the new PhysiBoSS functionalities. M.R., A.C., L.C., V.N. developed the models. E.B., P.M., L.C. and V.N. supervised the project. M.R., A.C., R.H., E.B., P.M., L.C., V.N. wrote and reviewed the manuscript. 

\section*{Funding}
M.R., L.C. and V.N. were partially funded by the PerMedCoE project which is part of the European Union's Horizon 2020 research and innovation program under the grant agreement n951773, and by the Certainty project which is part of the European Union's Horizon Europe research and innovation program under Grant agreement n101136379. M.R., L.C. and V.N. work also received funding from the Inserm amorçage project. R.H. and P.M. received funding from the National Science Foundation (Awards 1720625 and 2303695), the National Institutes of Health (U01-CA232137-01), and the Jayne Koskinas Ted Giovanis Foundation for Health and Policy. A.C, L.C. and V.N. were partly supported by ModICeD project from MIC ITMO 2020.

\section*{Acknowledgments}
We thank John Metzcar for testing PhysiBoSS implementation and Furkan Kurtoglu and Arnau Montagud for the fruitful discussion about PhysiBoSS mapping. 

\section*{Bibliography}
\bibliography{biblio}

\end{document}